\documentstyle[proceedings,psfig]{crckapb} 
\begin{opening}
\title{THE STAR-FORMATION HISTORY OF THE UNIVERSE}

\author{MICHAEL ROWAN-ROBINSON}
\institute{Astrophysics Group, Blackett Laboratory,\\ 
Imperial College of Science Technology and Medicine\\
           Prince Consort Road, London SW7 2BZ}
\end{opening}

\begin{document}

% The \begin{document} command comes after the \end{opening}
% command.

\section{Introduction} 

Many authors have attempted to model the star formation history of the universe
(eg Cole et al 1994, Pei and Fall 1995) or equivalently the evolution of the
starburst galaxy population (eg Franceschini et al 1991, 1994, 1997, Blain and Longair 1993,
Pearson and Rowan-Robinson 1996, Guiderdoni et al 1998, Dwek et al 1998, Blain et al 1999).
Most of these studies explore a small range of models designed to fit the available
observations. 
In the present paper I report the results of a parameterized approach to the star formation
history of the universe, which allows a large category of possible histories to be
explored and quantified. The parametrized models can be compared with a wide range
of source-count and background data at far infrared and submillimetre wavelengths to
narrow down the parameter space that the star formation history can occupy. 

Throughout this paper I use the most recent Bruzual and Charlot star formation
scenarios, with a Salpeter IMF with mass cutoffs at 0.1 and 100 $M_{\odot}$, for which
the relevant conversion factors are:\\
$L(H\alpha)/L{\odot} = 10^{7.60} \dot{\phi}_{*}$,  (1)\\
$L(60 \mu m)/L{\odot} = 10^{9.66} \epsilon \dot{\phi}_{*}$,  (2)\\
$L(2800 \AA)/L{\odot} = 10^{9.35} \dot{\phi}_{*}$,  (3)\\
$L(2000 \AA)/L{\odot} = 10^{9.50} \dot{\phi}_{*}$,  (4)\\
$ \dot{\phi}_{*} = 42  \dot{\phi}_{Z}$,  (5)

where $ \dot{\phi}_{*},  \dot{\phi}_{Z}$, are the star and metal formation rates
in solar masses per year per cubic Mpc and $\epsilon$ is the fraction of opt-uv light
absorbed by dust (Rowan-Robinson et al 1997).
A Hubble constant of 50 $km/s/Mpc$ is used throughout, and $\Omega_o$ = 1 (unless otherwise stated).

\section{Star formation rate derived from ultraviolet data }
I first estimate the star formation history at uv wavelengths, with no correction for the
effects of dust.  For the local, z = 0, star formation rate, I have gone 
back to the H$\alpha$ data of Gallego et al (1995) in order to remove the reddening correction they 
have applied.  I find a mean reddening correction of 4.4, to give a value
for $\dot{\phi}_{*}$ of $10^{-2.76 \pm 0.20} M_{\odot}/yr/Mpc^{3}$.  
I have also estimated the star-formation rate from a 
sample of 170 bright (B $\leq$ 13.8) nearby (V $<$ 5000 km$/$s), optically selected galaxies 
for which there
are both 60 $\mu$m (S(60) $\ge$ 0.6 Jy) and large-beam 2000 \AA\/ observations
(Buat et al 1987, Kinney et al 1993, Deharveng et al 1994, Meurer et al 1995). 
The 60 $\mu$m and B-band luminosity functions can be used to 
verify that the sample is
representative of all galaxies and allow the effective area surveyed to be estimated.
The 2000 \AA\/ luminosity density is found to be  $10^{18.42} W Mpc^{-3}$ and the 
corresponding star formation rate is $10^{-2.48 \pm 0.06} M_{\odot} yr^{-1} Mpc^{-3}$.  This is
consistent with the revised estimate given above from the Gallego et al $H\alpha$ data
and with the estimate given by Lilly et al (1996) extrapolated from B-band data,
$10^{-2.42} M_{\odot} yr^{-1} Mpc^{-3}$, but is a factor of 3 lower than the value
used by Madau et al (1996).

\begin{figure}
\centerline{\psfig{file=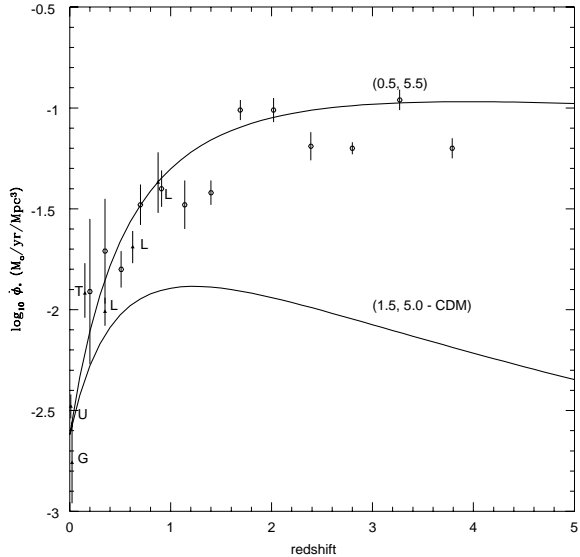,angle=0,width=8cm}}
\caption{Star formation history derived from ultraviolet data, with no
correction for effects of dust.  Data points are from Gallego et al (1995),
 (G), Treyer et al (1998) (T), Lilly et al (1996)
(L), from the sample of bright local galaxies (section 2) at z = 0.01 (U),  and
from the analysis of photometric redshifts of HDF galaxies (open circles).  
The models shown are chosen to fit the CDM 
predictions of Cole et al
(1996) (P,Q) = (1.5, 5.), and to account for the far infrared and submm counts,
 (P,Q) = (0.5, 5.5). 
}
\end{figure}

\begin{figure}
\centerline{\psfig{file=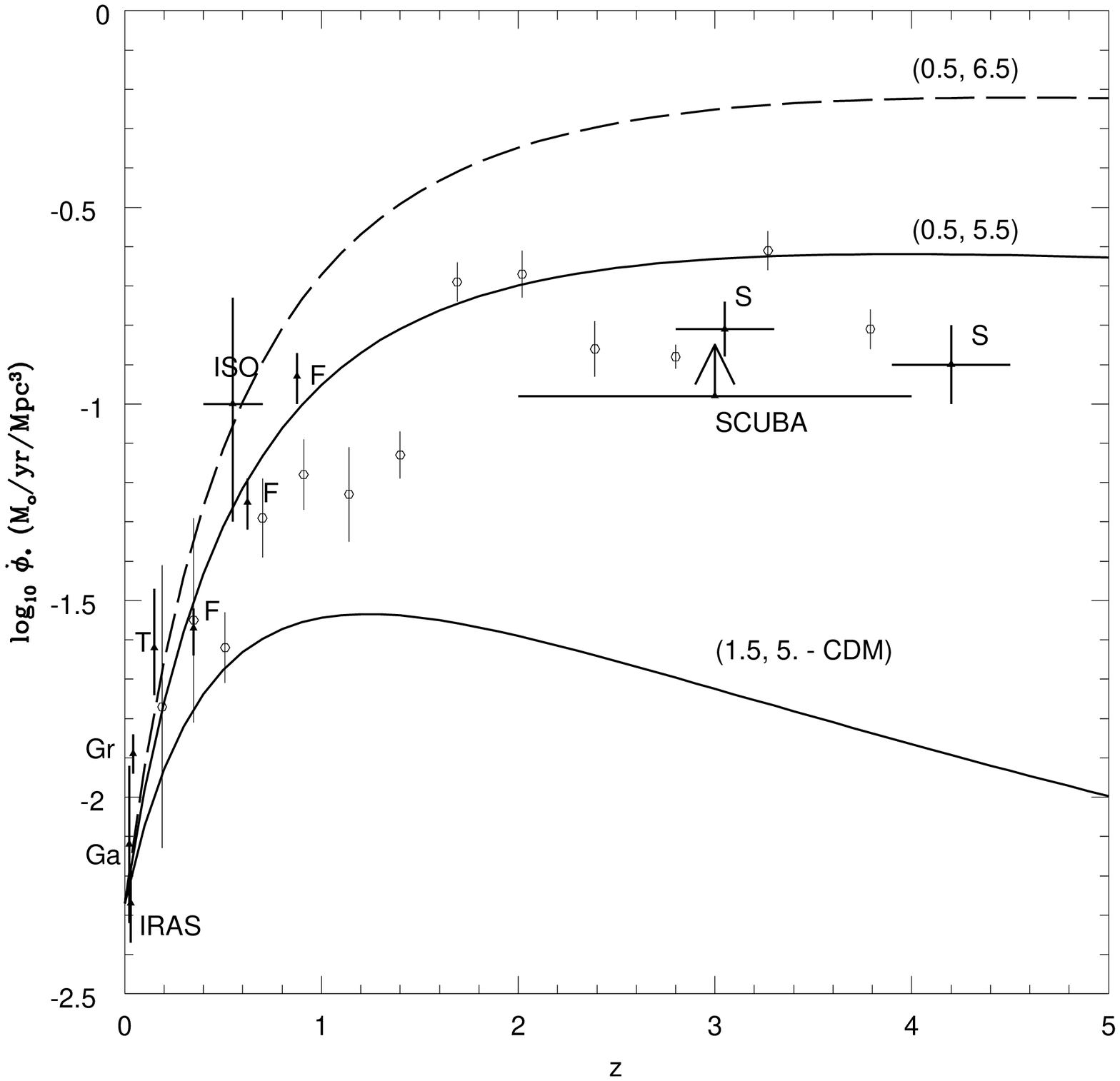,angle=0,width=8cm}}
\caption{Star formation history derived from infrared (IRAS: Saunders et al 1990, ISO: 
Rowan-Robinson et al 1997 , F: Flores et al 1997 ), submillimeter
(SCUBA, 850 $\mu$m: Hughes et al 1998), and from ultraviolet 
data with
correction for effects of dust, from Gallego et al (1995)
 (Ga), Treyer et al (1998) (T),
 both corrected by a factor 2 for dust extinction, Gronwall (1998) (Gr) and Steidel
et al (1999) (S) and from the 
analysis of photometric redshifts of HDF galaxies 
assuming E(B-V) =0.1 (open circles).  
}
\end{figure}

Madau et al (1996) have given estimates for the star formation rate at z = 2 - 4.5
derived from analysis of Lyman drop-out galaxies in the Hubble Deep Field.
This analysis appeared to demonstrate that the star formation drops off
steeply at z $>$ 2.  
Connolly et al (1997) have used photometric redshifts for the brighter HDF galaxies
to derive the star formation rate at z = 0.5 - 2.  Their method involves the
use of J, H and K data in addition to the HST U,B,V,I data to determine the 
redshifts.  Although this certainly greatly improves the reliability of the
photometric redshifts, it restricts the analysis to 
$< 10 \%$ of the galaxies detected in the HDF by HST.

To try to extend the results of Connolly et al to a wider range of redshift, and to test
the robustness of the conclusions of Madau et al (1996), I have used the
photometric redshift method of Mobasher et al (1996) to analyze a much larger
sample of HDF galaxies, namely all those galaxies with $I_{AB} \leq$ 29.0 from
the catalogue of Williams et al (1996) detected
in at least the I and V bands, a total of 2438 galaxies once duplicates have
been eliminated.  This method gives redshifts accurate to 13 $\%$ in (1+z),
although there is obviously a possibility of aliasing for galaxies detected
in only 2 bands (22$\%$ of the sample). 
For these 2438 HDF galaxies I have estimated the luminosity functions of the galaxies at 
rest-frame 2800 \AA\/ in bins
of 0.05 in $log_{10}(1+z)$, for z $\leq$ 4.  
  The resulting star formation history
is shown in Fig 1.   
Agreement with the results of Lilly et al (1996), Treyer et al
(1997) and Connolly et al (1997) is excellent over the range z = 0.2-2.5.  
However the derived star formation
rates at higher redshifts are considerably higher than those of Madau et al (1996)
and shows no sign of a decline in star formation rate beyond z = 1.

\section{Star formation history from infrared or submillimetre data, or from 
ultraviolet data with correction for the effects of dust}

Figure 2 shows the star formation
history derived either from ultraviolet or H$\alpha$ data with correction for reddening,
or from far infrared or submillimetre data under the assumption that most ultraviolet and 
visible light
from star forming regions is absorbed by dust (i.e. $\epsilon \simeq$ 1).

In the case of the HDF data the extinction is derived using the grain properties of
Rowan-Robinson (1992) assuming a (conservative) extinction corresponding to E(B-V) = 0.1 
at the rest-frame wavelength correponding to $\lambda_{obs} = 6000$ \AA\/, which gives correction
factors ranging from 1.5 at z =0.2 to 2.5 at z = 3.8.  Most estimates of correction factors to 
be applied to high redshift star-forming galaxies range from 2-7 (Pettini et al 1997, 
Meurer et al 1998,
Calzetti 1998, Steidel et al 1999).  Note the good agreement of my values
derived from the HDF with those of Steidel et al (1999) from surveys for Lyman break galaxies.

The ISO-HDF estimates of Rowan-Robinson et al (1997) for the redshift range 0.4-0.7 have been 
corrected for the fact that one of the 5 galaxies in this range is no longer thought to be a starburst 
(source 3 in Table 1 of Rowan-Robinson et al).  This has an almost negligible effect on the total
star formation rate. Also shown are the star formation rates derived by Hammer and Flores (1998)
 from ISO observations of a CFRS field.   
    
The shape of the observed star formation history is similar to that derived from the uv 
(Fig 1), but the values are a factor 2-3 higher at most redshifts, implying $\epsilon$ =
2/3-3/4.  For comparison Calzetti (1998) gives a range for $\epsilon$ of 1/2 to
2/3 for starburst galaxies.

\section{Parametrized approach to star formation history}

To study what constraints on the star formation history can be derived from source-counts
and background intensity measurements at far infrared and submm wavelengths, I present here
 a parameterized approach to the problem, investigating a wide
range of possible star formation histories.

The constraints we have on the star formation rate, $\dot{\phi_{*}}$(t) are that: 
(i) it is zero for t = 0, (ii) it is finite at $t = t_{o}$,
(iii) it increases with z out to at least z = 1 ( and from (i) must eventually decrease
at high z).

A simple mathematical form consistent with these constraints is

\medskip

$\dot{\phi_{*}}(t)/\dot{\phi_{*}}(t_{o})$ = exp Q(1-$(t/t_{o}))$ $(t/t_{o})^P$  (6)

\medskip

where P and Q are parameters 
( P $>$ 0 to satisfy (i), Q $>$ 0 to satisfy (iii)).  I assume that $\dot{\phi_{*}}(t)$ = 0
for z $>$ 10.

Equation (6) provides a simple but versatile parameterization of the star formation history, 
capable of reproducing most physically realistic, single-population scenarios.    

Given an assumed (P,Q) I then determine the 60 $\mu$m luminosity function, using the
IRAS 1.2 Jy sample (Fisher et al 1995).  I fit this with the form assumed by Saunders et al (1990)

\medskip
$\eta(L) = C_{*} (L/L_{*})^{1-\alpha} e^{-0.5[log_{10}(1 + L/L_{*})/\sigma]^{2}}$   (7)

\medskip
and find that the luminosity function parameters can be well approximated as follows:

$\sigma$ = 0.765-0.04 W

$log_{10}L_{*}$ = 8.42 + 0.07 W - 2 $log_{10} (H_{o}/100)$

where  W = 0.825 Q - P,
and I have assumed fixed values for   $\alpha$ = 1.09, $C_{*}$ = 0.027 $(H_{o}/100)^{3}$.
It is not clear that previous studies have correctly taken account of the need to change
the 60 $\mu$m luminosity function as the rate of evolution is varied.  The study of
Guiderdoni et al (1998) explicitly violates the known constraints on the 60 $\mu$m
luminosity function at the high luminosity end and as a result the models predict far too many
high redshift galaxies at a flux-limit of 0.2 Jy at 60 $\mu$m, where substantial redshift surveys have
already taken place.  

To transform this 60 $\mu$m luminosity function to other wavelengths I assume 
that the spectral energy distributions
of galaxies are a mixture of two components, a starburst component and a 'cirrus'
component (cf Rowan-Robinson and Crawford 1989).  I have used the latest predictions
for infrared seds of these two components by Efstathiou et al (1999).  

There is no explicit allowance for emission from AGN in the models, though it is known that
at low fir luminosities, about 10 $\%$ of galaxies contain AGN, with the proportion increasing
towards higher luminosities.  Provided the far infrared component peaking at 60 $\mu$m which I have
identified as the 'starburst' component is indeed powered by a starburst
(cf Rowan-Robinson and Crawford 1989, Rigopoulou et al 1996, Lutz et al 1998), 
it is irrelevant for estimates of the star formation rate
whether the galaxy also contains an AGN.  I am implicitly assuming that any correlation between far
infrared luminosity and the optical/uv/X-ray luminosity from an accreting black hole
is caused by a common feeding mechanism.  

I can now predict the counts and background intensity at any wavelength and by
comparing with observed values, constrain loci in the P-Q plane.  Fig 3 shows
a number of such loci for the case of pure luminosity
evolution in an Einstein de Sitter model ($\Omega$ = 1).  The uncertainties on
these loci are typically $\pm$0.5 in Q, except for the 60 $\mu$m counts, where
the uncertainty is much broader ($\pm$2 in Q), with rather weak dependence on P
for P $>$ 0.  The locus shown for N60 is in fact the upper end of the
1-$\sigma$ range for the number of sources per ster at 60 $\mu$m at 0.25 mJy.  
Figure 4,5 shows integral counts at 850 and 60  $\mu$m.  
Figure 6 shows the spectrum of the integrated background radiation,
compared with selected models.  Because of the simplistic nature of the assumed optical seds,
I have not attempted to show optical counts, but it is of interest to try to get roughly
the correct balance of optical and far infrared/submm background radiation.  

We see that models consistent with the far infrared and
submillimeter counts can be found, eg (P,Q) = (0.5,5.5), and that such models also are consistent
with the observed 850 $\mu$m background.  However stronger evolution is required to fit
the 140-350 $\mu$m background intensity, eg (P,Q) = (0.5,6.5), and such models are inconsistent
with the observed counts.  The CDM model, which can be approximated by
 (P,Q) = (1.5, 5.0), is consistent with
 the 60 $\mu$m counts but nothing else. 
 
An important constraint on the models is that the total mass of stars produced in galaxies
should be greater than or equal to the mass of stars observed, $\Omega_{*} \ge 0.006 \pm 0.0018 
(H_{o}/50)^{-1}$ 
(Lanzetta et al 1996), and that it should be less than the total density of baryons
in the universe, $\Omega_* \le 0.05 \pm 0.01 (H_{o}/50)^{-2}$ (Walker et al 1991).  
 The models which fit the counts ( (P,Q) = (0.5, 5.5) ) 
and 140-350 $\mu$m background ( (P,Q) = (0.5, 6.5) ), give  $\Omega_*$ = 0.015 and 0.032 
respectively.

\begin{figure}
\centerline{\psfig{file=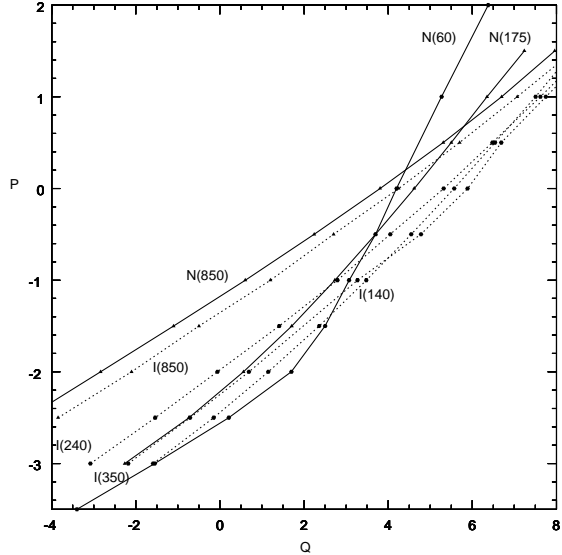,angle=0,width=8cm}}
\caption{
P-Q diagram, with loci for models fitting 60, 175 and 850 $\mu$m counts (solid curves) and 
140, 240, 350 and 850 $\mu$m background intensity (dotted curves).  Counts
loci correspond to  N60(0.25Jy) = 2.0, N175(0.1Jy) = 39 (Kawara et al 1998),
log N850(4mJy) = 900 per sq deg (Hughes et al 1998).  Background loci correspond 
to $\nu I_{\nu}$(140) = 12.5,
 $\nu I_{\nu}$(240) = 11, $\nu I_{\nu}$(350) = 6.0 and $\nu I_{\nu}$(850) = 0.5 $nW/m^2/sr$ 
(Fixsen et al 1998).}  
\end{figure}

\begin{figure}
\centerline{\psfig{file=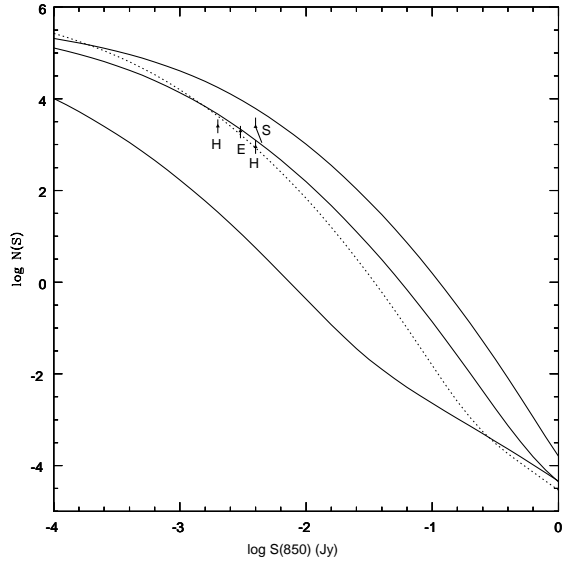,angle=0,width=8cm}}
\caption{
Integral source counts at 850 $\mu$m.   Data are from Hughes et al (1998), Eales et al (1999),
 Smail et al (1997).  The 3 models shown are, from bottom at
faint fluxes, for $\Omega_o = 1$ and (P,Q) = (1.5, 5.0), 
(0.5, 5.5) and (0.5, 6.5) (solid curves) and for $\Omega_o = 0.3$, (P,Q) = 
(1.5, 8.2) (dotted curve).}
\end{figure}

\begin{figure}
\centerline{\psfig{file=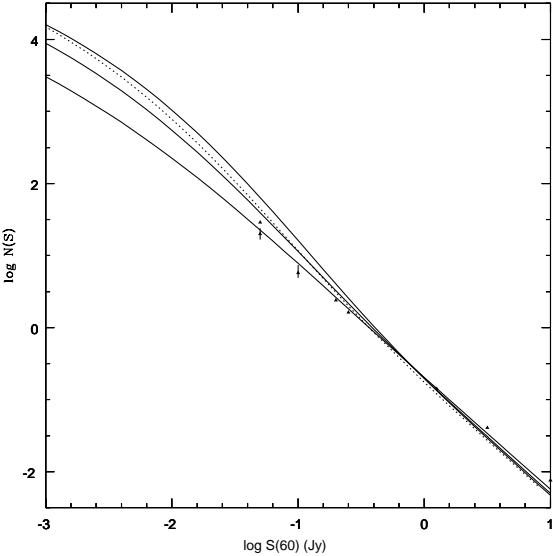,angle=0,width=8cm}}
\caption{
Source counts at 60 $\mu$m.   Data are from Lonsdale et al (1990) (at 0.2-10 Jy), Hacking and Houck 
(1987) (at 50-100 mJy),
Gregorich et al (1995) (higher point at 50 mJy).  Models as in Fig 4.}
\end{figure}

\begin{figure}
\centerline{\psfig{file=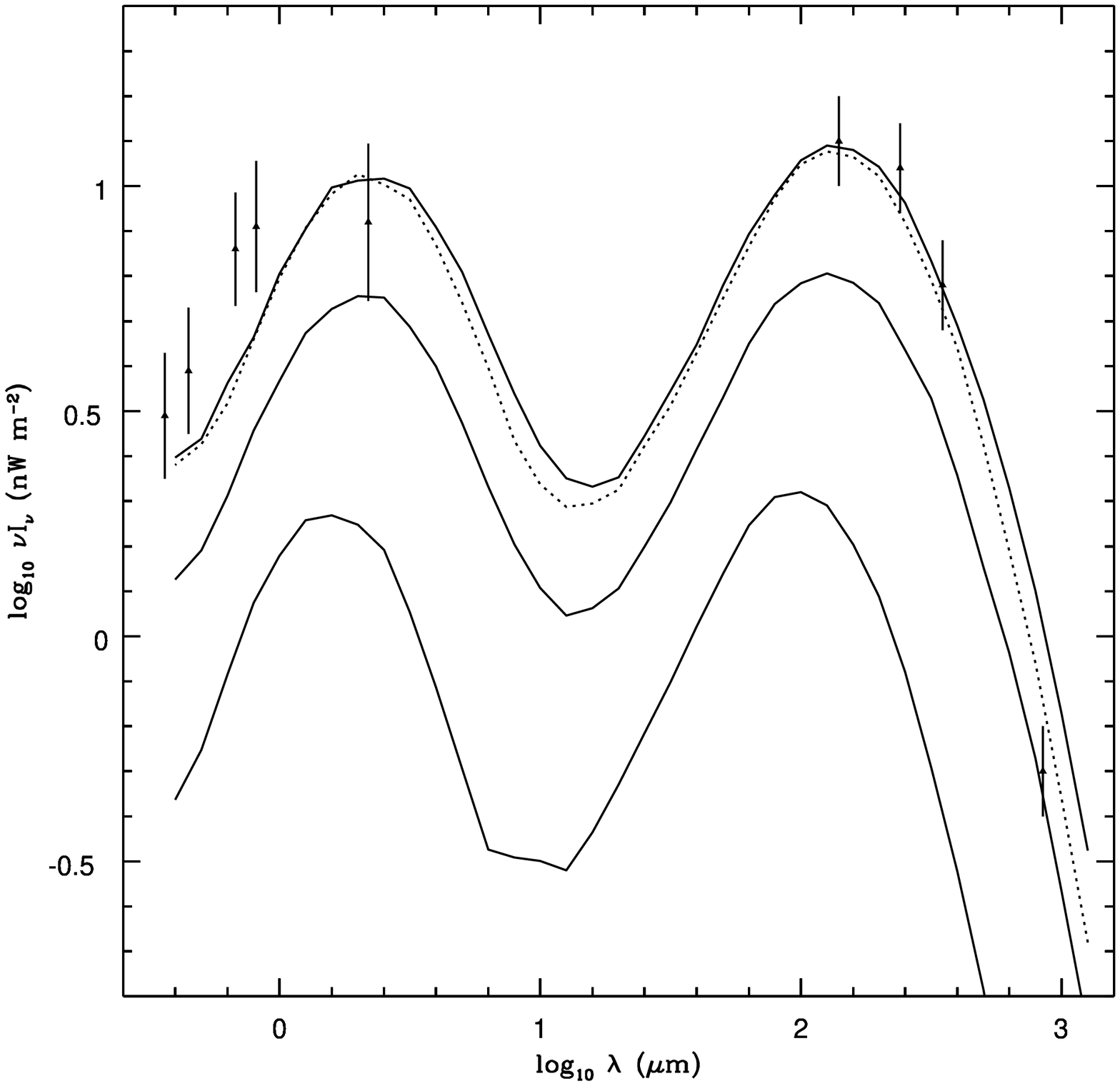,angle=0,width=8cm}}
\caption{
Predicted spectrum of integrated background for same models as Fig 4.
Data from Fixsen et al (1998)
(far ir and submm), Pozzetti et al (1998) (opt and uv).}
\end{figure}

We might expect that the cosmological model could have a significant effect on the
relationship between predicted counts and predicted background intensity, since
the latter is sensitive to how the volume element and look-back time change with
redshift.

\begin{figure}
\centerline{\psfig{file=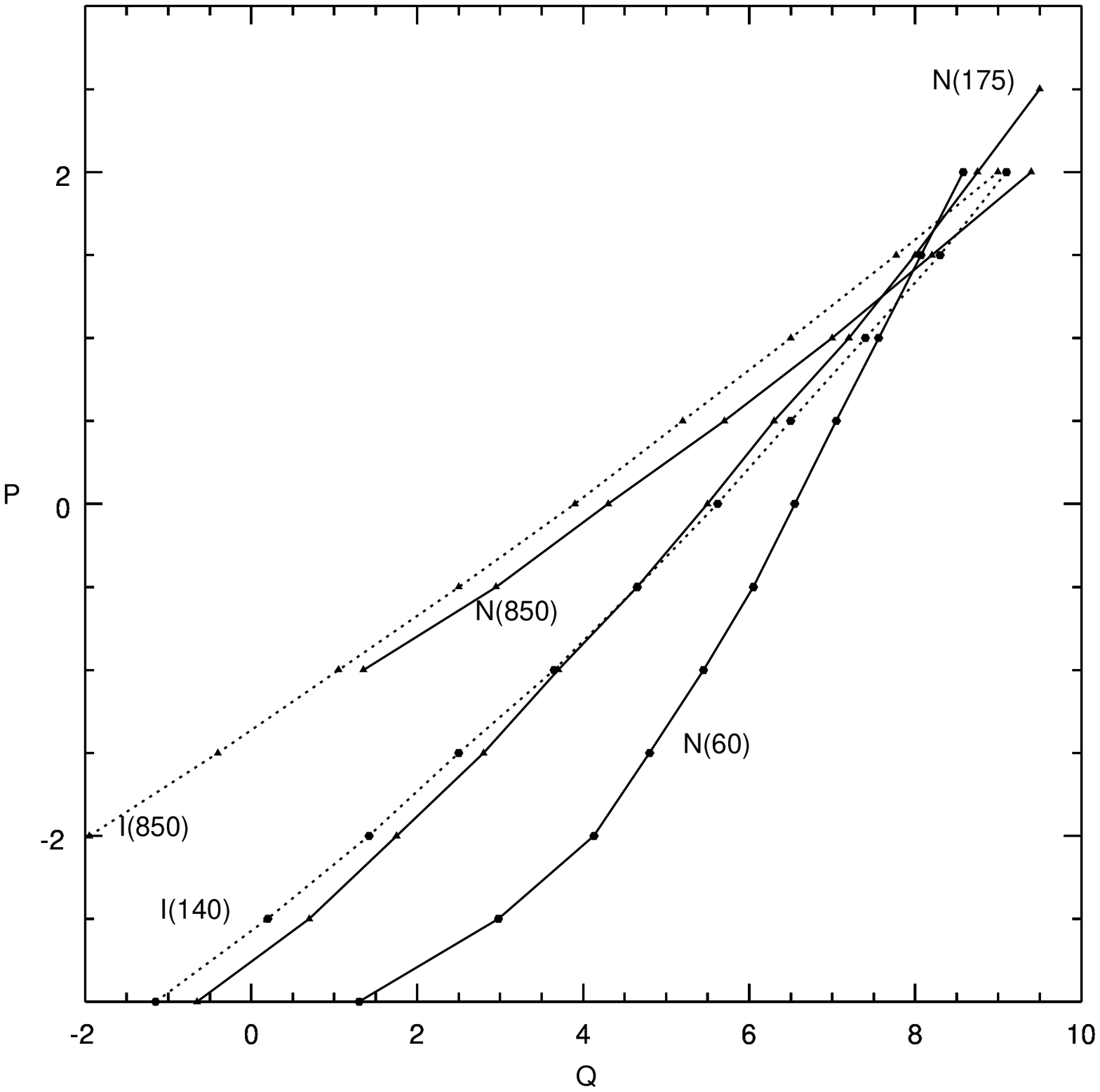,angle=0,width=9cm}}
\caption{
P-Q diagram  for $\Omega_o = 0.3$ model (section 7), with loci for models fitting 60, 175 and 
850 $\mu$m counts (solid curves) and 
140 and 850 $\mu$m background intensity (dotted curves).}
\end{figure}

To test this we have explored models with $\Lambda$ = 0, for which all the required 
formulae are analytic.  
Figure 7 shows the P-Q diagram for a model with $\Omega_{o}$ = 0.3, with loci similar to 
those of Fig 3 plotted.  In this cosmological model there does appear to be a single star 
formation history which is
consistent with both source-count and background radiation data, namely one with P = 1.5,
Q = 8.2.  Predicted counts and background spectrum are shown for this model as the
dotted curve in Figs 4-6.  

Thus within the framework of the types of star formation history considered in section 5,
there is a clear sensitivity to cosmological parameters, with  $\Omega_{o}$ = 0.3 preferred
to  $\Omega_{o}$ = 1.  This not to say that more complex assumptions about the star formation history,
for example the inclusion of a dust-enshrouded population not represented at low redshift as in
Franceschini et al (1997),
might not be consistent with  $\Omega_{o}$ = 1.
The full dependence on cosmological parameters will be explored in later work.

%\section*{Acknowledgements} 


\begin{thebibliography}{}

%\bibitem[]{} Aussel H., Cesarsky C.J., Elbaz D., Starck J.L., 1998, AA (submitted), astro-ph/9810044

%\bibitem[]{} Barger A.J., Cowie L.L., Sanders D.B., Fulton E., Taniguchi Y., Sato Y., Kawara K., 
%Okuda H., 1998, Nat 394, 248

%\bibitem[]{} Baugh C., Cole S., Frenk C., Lacey C., 1998, ApJ 498, 504

%\bibitem[]{} Bertin E., Dennefeld M., Moshir M., 1997, AA 323, 685

%\bibitem[]{} Bruzual A.G., and Charlot, S., 1993, ApJ 405, 538

\bibitem[]{} Blain A.W., Longair M.S., 1993, MN 264, 509

\bibitem[]{} Blain A.W., Smail I., Ivison R.J., Kneib J.-P., 1999, MN 302, 632

\bibitem[]{} Buat V., Donas J., Deharveng J.-M., 1987, AA 185, 33

\bibitem[]{} Calzetti F., 1998, in 'Dwarf Galaxies and Cosmology', ed. T.X.Thuan, C.Balkowski, V.Cayatte,
J. Tran Thanh Van (Editions Frontieres) astro-ph/9806083

%\bibitem[]{} Calzetti F., Heckman T.M., 1998, ApJ (in press), astro-ph/9811099

\bibitem[]{} Cole S., Aragon-Salamanca A., Frenk C.S., Navarro J.F., Zepf S.E., 1994, MN 271, 781

\bibitem[]{} Connolly A.Z., Szalay A.S., Dickinson M., Subbarao M.U., Brunner P.J., 1997, ApJ 486, L11

%\bibitem[]{} Cram L., 1998, ApJ (in press), astro-ph/9808228

\bibitem[]{} Deharveng J.-M., Sasseen T.P., Buat V., Lampton M., Wu X., 1994, AA 289, 715

%\bibitem[]{} Dickinson, M., 1998, in 'The Hubble Deep Field', ed. M.Livio, S.M.Fall and P.Madau (STScI 
%Symposium Series), astro-ph/9802064

\bibitem[]{} Dwek E., et al, 1998, ApJ 508, 106

\bibitem[]{} Eales S., Lilly S.J., Gear W.K., Dunne L., Bond J.R., Hammer F., Le Fevre O., Crampton D., 
1999, ApJL  (in press), astro-ph/9808040

\bibitem[]{} Efstathiou A., Rowan-Robinson M., Siebenmorgen R., 1999, MN (in press)

\bibitem[]{} Fisher K.B., Huchra J.P., Strauss M.A., Davis M., Yahil A., Schlegel D., 1995, ApJS 100, 69

\bibitem[]{} Fixsen D.J., Dwek E., Mather J.C., Bennett C.L., Shafer R.A., 1998, ApJ 508, 123 

\bibitem[]{} Franceschini A., Toffolatti L., Mazzei P., Danese L., De Zotti G., 1991, AAS 89, 285 

\bibitem[]{} Franceschini A.,  Mazzei, P.,  de Zotti, G.,  Danese, L., 1994, ApJ 427, 140

\bibitem[]{} Franceschini A., Silva L., Fasano G., Granato G.L., Bressan A., Arnouts S., Danese L., 
1998, ApJ 506, 600

\bibitem[]{} Gallego J., Zamorano J., Aragon-Salamanca A., Rego M., 1995, ApJ 455, L1

%\bibitem[]{} Glazebrook K., Blake C., Economou F., Lilly S., Colless M., 1998. MN (in press), 
%astro-ph/9808276

\bibitem[]{} Gregorich D.T., Neugebauer G., Soifer B.T., Gunn J.E., Herter T.L., 1995, AJ 110, 259

\bibitem[]{} Gronwall C., 1998, in 'Dwarf Galaxies and Cosmology', eds T.X.Thuan, C.Balkowski, 
V.Cayatte, J.Tran Thanh Van, (Editions Frontieres), astro-ph/9806240

\bibitem[]{} Guiderdoni B, Hivon E., Bouchet F.R., Maffei B., 1998, MN 295, 877

\bibitem[]{} Hacking P.B. and Houck J., 1987, ApJS 63, 311

\bibitem[]{} Hammer F., Flores H., 1998, 
in 'Dwarf Galaxies and Cosmology', eds T.X.Thuan, C.balkowski, V.Cayatte, J.Tran Thanh Van, 
(Editions Frontieres), astro-ph/9806184

\bibitem[]{} Hauser M.G., et al, 1998, ApJ 508,25

%\bibitem[]{} Holland W.S, 1998, Nat 392, 788

\bibitem[]{} Hughes D.H., et al, 1998, Nat 394, 241

\bibitem[]{} Kawara K., et al, 1998, AA (in press)

\bibitem[]{} Kinney A.L., Bohlin R.C., Calzetti D., Panagia N., Wyse R.F.G., 1993, ApJS 86, 5

%\bibitem[]{} Lanzetta K.M., Yahil A., Fernandez-Soto A., 1996, Nature 381, 759

%\bibitem[]{} Leech K.J., Penston M.V., Terlevich R., Lawrence A., Rowan-Robinson M., Crawford J., 
1989, MN 240, 349

\bibitem[]{} Lilly, S.J., Le Fevre, O., Hammer, F., Crampton, D., 1996, ApJ 460, L1

\bibitem[]{} Lonsdale C.J., Hacking P.B., Conrow T.P., Rowan-Robinson M., 1990, ApJ 358, 60

\bibitem[]{} Lutz D., Spoon H.W.W., Rigopoulou D., Moorwood A.F.M, Genzel R., 1998, 
ApJ 505, L103

\bibitem[]{} Madau, P., Ferguson, H.C., Dickinson, M.E., Giavalisco, M., Steidel, C.C., Fruchter, A., 
1996, MNRAS 283, 1388

%\bibitem[]{} Madau, P., 1997, in 'Star Formation Near and Far', (Un.iv. of Maryland) astro-ph/9612157

%\bibitem[]{} Madau, P., 1998, in 'The Hubble Deep Field', ed. M.Livio, S.M.Fall and P.Madau (STScI 
%Symposium Series) astro-ph/9709147

%\bibitem[]{} Madau P., Pozzetti L., Dickinson M., 1998, ApJ 498, 106

\bibitem[]{} Meurer G.R., Heckman T.M., Leitherer C., Kinney A., Robert C., Garnett D.R., 1995, 
AJ 110, 2665

%\bibitem[]{} Meurer G.R., Heckman T.M., Lehnert M.D., Leitherer C., Lowenthal J., 1997, AJ 114, 54

%\bibitem[]{} Meurer G.R., 1998, in 'Hubble Deep Field', ed.  M.Livio, S.M.Fall and P.Madau (STScI 
%Symposium Series) astro-ph/9708163

%\bibitem[]{} Meurer G.R., Heckman T.M., Calzetti D., 1998, in preparation

%\bibitem[]{} Mazzarella J.M., Balzano V.A., 1986, ApJS 62, 751

\bibitem[]{} Mobasher, B., Rowan-Robinson, M., Georgakakis, A., Eaton, N., 1996, MN 282, L7

%\bibitem[]{} Oliver, S., {\it et al.}, 1995, in 'Wide-Field Spectroscopy and the Distant Universe', eds.
%S.J.Maddox and A.Arogon-Salamanca (World Scientific) p.274

%\bibitem[]{} Oliver, S., Gruppioni C., Serjeant S., 1998, MN (submitted), astro-ph/9808260

\bibitem[]{} Pearson, C., Rowan-Robinson, M., 1996, MN 283, 174

\bibitem[]{} Pettini M., et al,
1997, in 'Cosmic Origins: Evolution of Galaxies, Stars, Planets and Life', eds Woodward C.E., 
et al, ASP Conf.Ser. (astro-ph/9708117)

%\bibitem[]{} Pettini M., Kellogg M., Steidel C.C., Dickinson M., Adelberger K.L.,, Giavalisco M., 
%1998, ApJ in press (astro-ph/9806219)

%\bibitem[]{} Pettini M., Steidel C.C., Kellogg M., Dickinson M., Adelberger KL., Giavalisco M., 1998, 
%in 'Dwarf Galaxies and Cosmology', eds T.X.Thuan, C.balkowski, V.Cayatte, J.Tran Thanh Van, 
%(Editions Frontieres)

\bibitem[]{} Pei, Y.C., and Fall, S.M., 1995, ApJ 454, 69

\bibitem[]{} Pozzetti L., Madau P., Ferguson H.C., Zamorani G., Bruzual G.A., 1998, MN 298,1133

\bibitem[]{} Puget, J.-L., Abergal A., Bernard J.-P., Boulanger F., Burton W.B., 
Desert F.-X., Hartmann D., 1966, AA 308, 5

\bibitem[]{} Rigopoulou D., Lawrence A., Rowan-Robinson M., 1996, MN 288, 1049

%\bibitem[]{} Rowan-Robinson, M., Helou, G., Walker, D., 1987, MN 227, 589

\bibitem[]{} Rowan-Robinson M., and Crawford J., 1989, MN 238, 523

\bibitem[]{} Rowan-Robinson, M., 1992, MN 258, 787

%\bibitem[]{} Rowan-Robinson M., 1995, MNRAS 272, 737

%\bibitem[]{} Rowan-Robinson, M., Efstathiou, A., 1993, MN 263, 675

%\bibitem[]{} Rowan-Robinson, M., Pearson, C., 1996, in 'Unveiling the Infrared Background',
%ed. E.Dwek (American Inst. of Physics) p.192

%\bibitem[]{} Rowan-Robinson M., Benn C.R., Lawrence A., McMahon R.G., Broadhurst T.J., 1993, MNRAS 263, 123

\bibitem[]{} Rowan-Robinson M. et al, 1997, MN 289, 490

\bibitem[]{} Saunders, W., Rowan-Robinson, M., Lawrence, A., Efstathiou, G., Kaiser, N.,
Frenk, C.S., 1990, MN 242, 318

\bibitem[]{} Schlegel D.J., Finkbeiner D.P., Davis M., 1998, ApJ 500, 525

\bibitem[]{} Smail I., Ivison R.J., Blain A.W., 1997, ApJL 490, L5 

%\bibitem[]{} Smail I., Ivison R.J., Blain A., Kneib J.-P., 1998, in 'After the dark ages: when galaxies 
%were young', (Univ. of Maryland), astro-ph/9810281

%\bibitem[]{} Somerville R.S., Primack J.R., 1998, Xth Rencontres de Blois, astro-ph/9811001

\bibitem[]{} Steidel C.C., Adelberger K.L., Giavalisco M., Dickinson M., Pettini M., 1999, 
ApJ (in press), astro-ph/9811399

%\bibitem[]{} Tresse L., Maddox S.J., 1998, ApJ 495, 691 

\bibitem[]{} Treyer M.A., Ellis R.S., Milliard B., Donas J., Bridges T.J., 1998, MN 300, 303

\bibitem[]{} Walker T.P., Steigman G., Schramm D.N., Olive K.A., Kang H.-S., 1991, ApJ 376, 51

\bibitem[]{} Williams R.E. et al, 1996, AJ 112, 1335

%\bibitem[]{} Wilner D.O., Wright M.C.W., 1997, ApJ 488, L67

\end{thebibliography}
\end{document}